\documentstyle[aps,floats,epsfig,psfig,psfrag,fleqn,eqsecnum]{revtex}

\DeclareFontFamily{U}{msb}{}
\DeclareFontShape{U}{msb}{m}{n}{
<5><6><7><8><9> gen *msbm <10><10.95><12><14.4><17.28><20.74><24.88>msbm10}{}
\DeclareSymbolFont{AMSb}{U}{msb}{m}{n}
\DeclareMathSymbol{\GGG}{\mathbin}{AMSb}{'107}
\DeclareMathSymbol{\KKK}{\mathbin}{AMSb}{'113}
\DeclareMathSymbol{\MMM}{\mathbin}{AMSb}{'115}
\DeclareMathSymbol{\PPP}{\mathbin}{AMSb}{'120}
\DeclareMathSymbol{\RRR}{\mathbin}{AMSb}{'122}
\DeclareMathSymbol{\SSS}{\mathbin}{AMSb}{'123}
\DeclareMathSymbol{\ZZZZ}{\mathbin}{AMSb}{'132}

\newcommand{\ch}[1]{\chi\rule[-1ex]{0mm}{2ex}_{#1}}
\newcommand{\gex}[1]{\left\langle #1 \right\rangle}
\newcommand{\rai}{\rightarrow \infty}

\newcommand{\phb}{\overline{\phi}}

\newcommand{\GT}{\widetilde{G}}

\newcommand{\XX}{{\bf r}}
\newcommand{\KK}{{\bf k}}
\newcommand{\EX}{{\sf EX}}
\newcommand{\DM}{{\sf DM}}
\newcommand{\GF}{{\sf GF}}
\newcommand{\SA}{{\sf SA}}

\def\DIRECTORY{.}

\begin{document}
\setcounter{page}{0}
\title{
Quantitative Analysis of Experimental and Synthetic Microstructures 
for Sedimentary Rock
}

\author{B. Biswa$\mbox{\rm l}^{1,2}$, C. Manwar$\mbox{\rm t}^{1}$,
R. Hilfe$\mbox{\rm r}^{1,3}$, S. Bakk$\mbox{\rm e}^{4}$ and 
P.E. {\O}re$\mbox{\rm n}^{4}$}

\address{
$\mbox{ }^1$ICA-1, Universit{\"a}t Stuttgart,
Pfaffenwaldring 27, 70569 Stuttgart\\
$\mbox{ }^2$Department of Physics \& Electronics,
Sri Venkateswara College, \\
University of Delhi, New Delhi - 110 021, India\\
$\mbox{ }^3$Institut f{\"u}r Physik,
Universit{\"a}t Mainz,
55099 Mainz, Germany\\
$\mbox{ }^4$ Statoil Research Center, 7004 Trondheim, Norway}

\maketitle

\thispagestyle{empty}

\begin{abstract}
A quantitative comparison between the experimental microstructure of a
sedimentary rock and three theoretical models for the same rock is
presented.  The microstructure of the rock sample (Fontainebleau
sandstone) was obtained by microtomography.  Two of the models are
stochastic models based on correlation function reconstruction, and
one model is based on sedimentation, compaction and diagenesis
combined with input from petrographic analysis.  The porosity of all
models closely match that of the experimental sample and two models
have also the same two point correlation function as the experimental
sample.  We compute quantitative differences and similarities between
the various microstructures by a method based on local porosity
theory.  Differences are found in the degree of anisotropy, and in
fluctuations of porosity and connectivity.  The stochastic models
differ strongly from the real sandstone in their connectivity
properties, and hence need further refinement when used to model
transport.
\end{abstract}

\vspace{.2cm}
\begin{tabbing}
PACS: \= 61.43.G \hspace*{2ex}\= (Porous materials; structure),\\
\> 81.05.Rm \> (Porous materials; granular materials),\\
\> 47.55.Mh \> (Flows through porous media)
\end{tabbing}

\vspace*{1cm}
{\em to appear in} : Physica A (1999), in print

\newpage

\section{Introduction}

A quantitative prediction of fluid flow, sound propagation, or
chemical transport in strongly correlated disordered media, such as
sedimentary rock, frequently employs representative microscopic models
of the microstructure as input.  A large number of microscopic models
have been proposed over the years to represent the microstructure of
porous media
\cite{fat56a,sch74,CD77,zim82,RS85,JA87,SK87,oxa91,adl92,BT93,sah95,FJ95,JC96,AAMHSS97,OBA98}.

Microscopic models do not reproduce the exact microstructure of the
medium at hand, but are based on the idea that the experimental sample
is a representative realization drawn from a statistical ensemble of
similar microstructures.  Hence it is necessary to have methods for
distinguishing microstructures quantitatively
\cite{hil95d,hil96g,hil92f,sic97}.
This is particularly important for attempts to generate porous
microstructures in an automatic computerized process
\cite{qui84,adl92,YT98a,YT98b,BT93}.

Despite the generality of the problem sketched above our discussion
will be focussed on fluid flow through sedimentary rocks.  In
particular we will discuss Fontainebleau sandstone.  This model system
has (together with Berea sandstone) acquired the status of a reference
standard for modeling and analysing sedimentary rocks
\cite{BZ85,BCZ87,adl92,YT98a,TSA93}. 

General geometric characterization methods traditionally include
porosities, specific surface areas, and sometimes correlation
functions \cite{sch74,adl92,ste85,dul92,BO97}.  Recently a more
refined, quantitative characterization for general stochastic
microstructures was based on local porosity theory (LPT)
\cite{hil92a,hil95d,hil91d,hil92b,hil92f,hil93b,hil93c,hil94b,hil98a}.
LPT is currently the most general geometric 
characterization method because it contains  as a special case 
also the characterization through correlation functions
(see \cite{hil95d} for details). 

Local porosity theory is used in this paper to distinguish
quantitatively various models for Fontainebleau sandstone.  More
precisely, the objective of this work is to give a quantitative
comparison of four microstructures.  One of them is an experimental
sample of Fontainebleau sandstone, while three of the microstructures
are synthetic samples from computer simulation models for
Fontainebleau sandstone.  One of the models is a sedimentation and
diagenesis model that tries to mimick the formation of sandstone
through deposition and cementation of spherical grains.  Two purely
stochastic models generate random realizations of microstructures with
prescribed porosity and correlation function.  The first of these is
based on Fourier space filtering of Gaussian random fields, and the
second is based on a simulated annealing algorithm.

In section \ref{lpt} we introduce and define the geometrical
quantities that will be used to distinguish the microstructures.  In
section \ref{samples} we present the four microstructures, their
generation and characterization in terms of the generation procedure.
In section \ref{results} we present the results and discuss the
differences between the four microstructures.

\section{Measured Quantities}
\label{lpt}

\subsection {Porosity and Correlation Functions}

Consider a rock sample occupying a subset $\SSS\subset\RRR^d$ of the
physical space ($d=3$ in the following).  The sample $\SSS$ contains
two disjoint subsets $\SSS=\PPP\cup\MMM$ with $\PPP\cap\MMM=\emptyset$
where $\PPP$ is the pore space and $\MMM$ is the rock or mineral
matrix and $\emptyset$ is the empty set.  The porosity $\phi(\SSS)$ of
such a two component porous medium is defined as the ratio $\phi(\SSS)
= V(\PPP)/V(\SSS)$ which gives the volume fraction of pore space.
Here $V(\PPP)$ denotes the volume of the pore space, and $V(\SSS)$ is
the total sample volume.

For the sample data analysed here the set $\SSS$ is a rectangular
parallelepiped whose sidelengths are $M_1,M_2$ and $M_3$ in units of
the lattice constant $a$ (resolution) of a simple cubic lattice.  Thus
the sample is represented in practice as the subset $\SSS =
[1,M_1]\times[1,M_2]\times[1,M_3]\subset \ZZZZ^3$ of an infinite cubic
lattice.  $\ZZZZ$ denotes the set of integers, and
$[1,M_i]\subset\ZZZZ$ are intervals.  The position vectors
$\XX_i=\XX_{i_1...i_d}=(a i_1,...,a i_d)$ with integers $1\leq i_j
\leq M_j$ are used to label the lattice points, and $\XX_i$ is a
shorthand notation for $\XX_{i_1...i_d}$.  A configuration (or
microstructure) ${\cal Z}$ of a $2$-component medium is then given as
\begin{equation}
{\cal Z}=(Z_1,...,Z_N)=(\ch{\PPP}(\XX_1),...,\ch{\PPP}(\XX_N))
\end{equation}
where $N=M_1M_2M_3$, and
\begin{equation}
\ch{\GGG}(\XX) = \left\{
\begin{array}{r@{\quad:\quad}l}
1 & \mbox{for}\quad \XX\in\GGG \\
0 & \mbox{for}\quad \XX\notin\GGG
\end{array}
\right .
\label{charfunc}
\end{equation}
is the characteristic (or indicator) function of a set $\GGG$ that
indicates when a point is inside or outside of $\GGG$.  A stochastic
medium is defined through the discrete probability density
\begin{equation}
p(z_1,...,z_N) = \mbox{Prob}\{(Z_1=z_1)\wedge...\wedge(Z_N=z_N)\}
\end{equation}
where $z_i\in\{0,1\}$. Expectation values of functions $f({\cal
Z})=f(z_1,...,z_N)$ are defined as
\begin{equation}
\langle f(z_1,...,z_N) \rangle = \sum_{z_1=0}^1...\sum_{z_N=0}^1
f(z_1,...,z_N)p(z_1,...,z_N)
\end{equation}
where the summations run over all configurations.  If the medium is
statistically homogeneous (stationary) then the average porosity is
given as
\begin{equation}
\gex{\phi}= \mbox{Prob}\{\XX_0\in\PPP\} = \gex{\ch{\PPP}(\XX_0)}
\label{avgporosity}
\end{equation}
where $\XX_0$ is an arbitrary lattice site.  If the medium is also
ergodic then the limit
\begin{equation}
\lim_{N\rai}\phi(\SSS)= \gex{\phi}
\end{equation}
exists.  There are, however, many subtleties associated with this
limit (see \cite{hil95d} for details).  Finally, we now define the
correlation function for a homogeneous medium as the expectation
\begin{equation}
  \label{eq:g_gen}
  G(\XX_0,\XX) = G(\XX-\XX_0) = 
  \frac{\gex{\ch{\PPP}(\XX_0)\ch{\PPP}(\XX)}-\gex{\phi}^2}
  {\gex{\phi}(1-\gex{\phi})} .
\end{equation}
If the medium is also isotropic $G(\XX)=G(|\XX|)=G(r)$.  Obviously
$G(0)=1$ and $G(\infty)=0$.

\subsection{Local Porosity Distributions}

The basic idea of local porosity theory is to measure geometric
observables within a bounded (compact) subset of the porous medium and
to collect these measurements into various histograms.  Let
$\KKK(\XX,L)$ denote a cube of sidelength $L$ centered at the lattice
vector $\XX$.  The set $\KKK(\XX,L)$ defines a measurement cell inside
of which local geometric properties such as porosity or specific
internal surface are measured \cite{hil91d}.  The local porosity in
this measurement cell $\KKK(\XX,L)$ is defined as
\begin{equation}
  \phi(\XX,L)=\frac{V(\PPP\cap\KKK(\XX,L))}{V(\KKK(\XX,L))}
  \label{lpd1}
\end{equation}
where $V(\GGG)$ is the volume of the set $\GGG\subset\RRR^d$.  The
local porosity distribution $\mu(\phi,L)$ is defined as
\begin{equation}
  \mu(\phi,L) = \frac{1}{m}\sum_\XX\delta(\phi-\phi(\XX,L))
  \label{lpd2}
\end{equation}
where $m$ is the number of placements of the measurement cell
$\KKK(\XX,L)$.  Ideally all measurement cells should be disjoint
\cite{hil91d}, but in practice this cannot be achieved because of poor
statistics.  The results presented below are obtained by placing
$\KKK(\XX,L)$ on all lattice sites $\XX$ which are at least a distance
$L/2$ from the boundary of $\SSS$, and hence in the following
\begin{equation}
  m = \prod^3_{i=1}(M_i-L+1) 
  \label{lpd3}
\end{equation}
will be used.  $\mu(\phi,L)$ is the empirical probability density
function (histogram) of local porosities.  Its support is the unit
interval.  In the following we denote averages with respect to
$\mu(\phi,L)$ by an overline.  Thus for a homogeneous and ergodic
medium
\begin{equation}
  \phb(L) = \int_0^1 \phi\mu(\phi,L)\;d\phi = \gex{\phi}
\end{equation}
is the expected local porosity.  In practice deviations from the last
equality may occur if the measurement cells are overlapping.  Figure
\ref{alp} below shows the average local porosity as function of $L$
for all four samples analyzed in this paper showing that deviations
can be as large as 0.5 percent.  The deviations may be partly
intrinsic and partly due to oversampling the central regions because
the measurement cells are overlapping.  Similarly the variance of
local porosities is found as \cite{hil95d}
\begin{eqnarray}
  \sigma^2 (L) = \overline{(\phi(L)-\phb(L))^2} & = &
  \int_0^1[\phi-\phb(L)]^2\mu(\phi,L)\;d\phi \nonumber\\
  & = & \frac{1}{L^3}\gex{\phi}(1-\gex{\phi})\left(1+
    \frac{2}{L^3}\sum_{{\XX_i,\XX_j\in\KKK(\XX_0,L)}\atop{i\neq j}}
    G(\XX_i-\XX_j)\right)
\label{variance}
\end{eqnarray}
where $\KKK(\XX_0,L)$ is any cubic measurement cell.

It is simple to determine $\mu(\phi,L)$ in the limits $L\rightarrow 0$
and $L\rightarrow \infty$ of small and large measurement cells.  For
small cells one finds generally \cite{hil91d,hil95d}
\begin{equation}
  \mu(\phi,L=0) = \phi(\SSS)\delta(\phi-1)+(1-\phi(\SSS))\delta(\phi)
  \label{lpd4}
\end{equation}
where $\phi(\SSS)$ is the sample porosity.  If the sample is
macroscopically homogeneous and ergodic then one expects
\begin{equation}
  \mu(\phi,L\rightarrow\infty) = \delta(\phi-\phi(\SSS))
  \label{lpd5}
\end{equation}
indicating that in both limits the geometrical information contained
in $\mu(\phi,L)$ consists of the single number $\phi(\SSS)$.  The
macroscopic limit, however, involves the question of macroscopic
heterogeneity versus macroscopic homogeneity (for more information see
\cite{hil95d}).  In any case, if eqs. (\ref{lpd4}) and (\ref{lpd5})
hold it follows that there exists a special length scale $L^*$ defined
as
\begin{equation}
  L^*=\min\{L : \mu(0,L)=\mu(1,L)=0\}
  \label{lpd6}
\end{equation}
at which the $\delta$-distributions at $\phi=0$ and $\phi=1$ both
vanish for the first time.

\subsection{Local Percolation Probabilities}

The local percolation probabilities characterize the connectivity of
measurement cells of a given local porosity.  Let
\begin{equation}
\Lambda_\alpha(\XX,L)=\left\{
\begin{array}{r@{\quad:\quad}l}
1 & {\rm if~}\KKK(\XX,L){\rm ~percolates~in~``}\alpha\mbox{\rm
''-direction}\\[6pt]
0 & {\rm otherwise}
\end{array}
\right.
\label{lpp1}
\end{equation}
be an indicator for percolation. What is meant by
``$\alpha$''-direction is summarized in Table~\ref{lpp}.  A cell
$\KKK(\XX,L)$ is called ``percolating in the $x$-direction'' if there
exists a path inside the set $\PPP\cap\KKK(\XX,L)$ connecting those
two faces of $\SSS$ that are vertical to the $x$-axis.  Similarly for
the other directions.  Thus $\Lambda_3=1$ indicates that the cell can
be traversed along all 3 directions, while $\Lambda_c=1$ indicates
that there exists at least one direction along which the block is
percolating.

The local percolation probability in the ``$\alpha$''-direction
is now defined through
\begin{equation}
\lambda_\alpha(\phi,L) = \frac{
\sum_\XX \Lambda_\alpha(\XX,L)\delta_{\phi\phi(\XX,L)}}
{\sum_\XX\delta_{\phi\phi(\XX,L)}} 
\label{lpp2}
\end{equation}
where $\delta_{\phi\phi(\XX,L)}=1$ if $\phi=\phi(\XX,L)$ and $0$
otherwise.  The local percolation probability $\lambda_\alpha(\phi,L)$
gives the fraction of measurement cells of sidelength $L$ with local
porosity $\phi$ that are percolating in the ``$\alpha$''-direction.

\subsection{Total Fraction of Percolating Cells}
The total fraction of all cells percolating along the
``$\alpha$''-direction is given by integration over all local
porosities as
\begin{equation}
  p_\alpha(L)=\int_0^1\mu(\phi,L)\lambda_\alpha(\phi,L)\;d\phi
  \label{pL1}
\end{equation}
This quantitiy provides an important characteristic for constructing
equivalent network models.  It gives the fraction of network elements
(bond, sites etc.) which have to be permeable in an equivalent
network.

\section{Despcription of Microstructures}
\label{samples}

\subsection{Experimental Sample of Fontainebleau Sandstone}

The experimental sample is a threedimensional microtomographic image
of Fontainebleau sandstone. This sandstone is a popular reference
standard because of its exceptional chemical, crystallographic and
microstructural simplicity
\cite{BZ85,BCZ87}.
Fontainebleau sandstone consists of monocrystalline quartz grains that
have been eroded for long periods before being deposited in dunes
along the shore during the Oligocene, i.e. roughly 30 million years
ago.  It is well sorted containing grains of around $200\mu$m in
diameter.  During its geological evolution, that is still not fully
understood, the sand was cemented by silica crytallizing around the
grains.  Fontainebleau sandstone exhibits intergranular porosity
ranging from $0.03$ to roughly $0.3$ \cite{BCZ87}.

The computer assisted microtomography was carried out on a micro-plug
drilled from a larger original core. The original core from which the
micro-plug was taken had a porosity of $0.1484$, a permability of $1.3
D$ and a formation factor of $22.1$.  The porosity $\phi(\SSS_\EX)$ of
our microtomographic data set is only 0.1355(see Table \ref{ovrvw}).
The difference between the porosity of the original core and that of
the final data set is due to the heterogeneity of the sandstone and to
the difference in sample size.  The experimental sample is referred to
as {\sf EX} in the following.  The pore space of the experimental
sample is visualized in Figure \ref{smplA}.

\subsection{Sedimentation, Compaction and Diagenesis Model}
The sedimentation, compaction and diagenesis model, abbreviated as
~\DM ~in the following, is obtained by numerically modelling the main
geological sandstone-forming processes \cite{BO97,OBA98}. Image
analysis of backscattered electron/cathodo-luminescence images of thin
sections provides input data such as porosity, grain size
distribution, a visual estimate of the degree of compaction, the
amount of quartz cement and clay contents and texture. The sandstone
modelling is carried out in three main steps: grain sedimentation,
compaction and diagenesis.  Here we give only a rough sketch of the
algorithms and refer the reader to \cite{BO97,OBA98} for a detailed
description.

Grain sedimentation commences with image analysis of thin sections.
The grain size distribution is measured using an erosion-dilation
algorithm.  Spherical grains with random diameters chosen from the
grain size distribution are dropped onto the grain bed and relaxed
into a potential energy minimum.  The sedimentation environment may be
low-energy (local minimum) or high-energy (global minimum).

Compaction reduces the bulk volume (and porosity) in response to
vertical stress from the overburden.  It is modelled here as a linear
process in which the vertical coordinate of every sandgrain is shifted
vertically downwards by an amount proportional to the original
vertical position.  The proportionality constant is called the
compaction factor.  Its value for our Fontainebleau sandstone is
estimated to be $0.1$ from thin section analysis.

In the diagenesis part only a subset of known diagenetical processes
are simulated, namely quartz cement overgrowth and precipitation of
authigenic clay on the free surface. Quartz cement overgrowth is
modeled by radially enlarging each grain.  If $R_0$ denotes the radius
of the originally deposited spherical grain, its new radius along the
direction $\XX$ from grain center is taken to be \cite{SK87,OBA98}
\begin{equation}
  R(\XX) = R_0 + \min(a\ell(\XX)^\gamma,\ell(\XX))
\end{equation}
where $\ell(\XX)$ is the distance between the surface of the original
spherical grain and the surface of its Voronoi ployhedron along the
direction $\XX$. The constant $a$ controls the amount of cement, and
the growth exponent $\gamma$ controls the type of cement overgrowth.
For $\gamma>0$ the cement grows preferentially into the pore bodies,
for $\gamma=0$ it grows concentrically, and for $\gamma<0$ quartz
cement grows towards the pore throats \cite{OBA98}. Authigenic clay
growth is simulated by precipitating clay voxels on the free mineral
surface.  The clay texture may be pore-lining or pore-filling or a
combination of the two.

For modeling the Fontainebleau sandstone we used a compaction factor
of $0.1$, and the cementation parameters $\gamma=-0.6$ and $a=2.9157$.
The resulting configuration of our sample ~\DM ~is displayed in Figure
\ref{smplB}.

\subsection{Gaussian Field Reconstruction Model}
\label{GF}
The Gaussian field ({\sf GF}) reconstruction model provides a random
pore space configuration in such a way that its correlation function
$G_\GF(\XX)$ equals a prescribed reference correlation function
$G_0(\XX)$.  In our case $G_0(\XX)=G_\EX(\XX)$ the reference is the
correlation function of the experimental sample described above.  The
method of Gaussian field reconstruction is well documented in the
literature \cite{qui84,AJQ90,adl92,YFKTA93}, and we shall only make a
few remarks that the reader may find of interest when implementing the
method.

Given the reference correlation function $G_\EX(\XX)$ and porosity
$\phi(\SSS_\EX)$ of the experimental sample the three main steps of
constructing the sample $\SSS_\GF$ with correlation function
$G_\GF(\XX)=G_\EX(\XX)$ are as follows:
\begin{enumerate}
\item
A standard Gaussian field $X(\XX)$ is generated which consists of
statistically independent Gaussian random variables $X\in\RRR$ at each
lattice point $\XX$.
\item
\label{l2}
The field $X(\XX)$ is first passed through a linear filter
which produces a correlated Gausssian field $Y(\XX)$ with
zero mean and unit variance.
The reference correlation function $G_\EX(\XX)$ and porosity 
$\phi(\SSS_\EX)$ enter into the mathematical construction 
of this linear filter.
\item
The correlated field $Y(\XX)$ is then passed through a nonlinear
discretization filter which produces the reconstructed sample $\SSS_\GF$.
\end{enumerate}

Details of these three main steps are documented in
Ref. \cite{qui84,AJQ90}.  However, in these traditional methods the
process described in step \ref{l2} is computationally difficult
because it requires the solution of a very large set of non-linear
equations. We have followed an alternate and computationally more
efficient method proposed in Ref. \cite{adl92} which uses Fourier
Transforms. For the sake of completeness we briefly describe
this. Later we shall discuss some of the difficulties experienced
while implementing this.

In the Fourier transform method the linear filter in step \ref{l2} is
defined in Fourier space through
\begin{equation}
  Y(\KK) = \alpha (G_Y(\KK))^{\frac{1}{2}}X(\KK),
\end{equation} 
where $M=M_1=M_2=M_3$ is the sidelength of a cubic sample, $\alpha =
M^{\frac{d}{2}}$ is the normalisation factor, and
\begin{equation}
  X(\KK) = \frac{1}{M^d}\sum_{\XX} X(\XX)e^{2\pi i\KK\cdot\XX}
\end{equation}
denotes the Fourier transform of $X(\XX)$. Similarly $Y(\KK)$ is the
Fourier transform of $Y(\XX)$, and $G_Y(\KK)$ is the Fourier transform
of the correlation function $G_Y(\XX)$.  $G_Y(\XX)$ has to be computed
by an inverse process from the correlation function $G_\EX(\XX)$ and
porosity of the experimental reference (details in \cite{adl92}).

It is important to note that the Gaussian field reconstruction
requires a large separation $\xi_\EX\ll N^{1/d}$ where $\xi_\EX$ is
the correlation length of the experimental reference, and
$N=M_1M_2M_3$ is the number of sites.  $\xi_\EX$ is defined as the
length such that $G_\EX(r)\approx 0$ for $r>\xi_\EX$.  If the
condition $\xi_\EX\ll N^{1/d}$ is violated then step \ref{l2} of the
reconstruction fails in the sense that the correlated Gaussian field
$Y(\XX)$ does not have zero mean and unit variance.  In such a
situation the filter $G_Y(\KK)$ will differ from the Fourier transform
of the correlation function of the $Y(\XX)$.  It is also difficult to
calculate $G_Y(r)$ accurately near $r=0$
\cite{adl92}.
This leads to a discrepancy at small $r$ between $G_\GF(r)$ and
$G_\EX(r)$.  The problem can be overcome by choosing large $M$ as we
verified in $d=1$ and $d=2$.  However, in $d=3$ very large $M$ also
demands prohibitively large memory.  In earlier work
\cite{AJQ90,adl92} the correlation function $G_\EX(\XX)$ was sampled
down to a lower resolution, and the reconstruction algorithm then
proceeded with such a rescaled correlation function.  This leads to a
reconstructed sample $\SSS_\GF$ which also has a lower resolution.
Such reconstructions have lower average connectivity compared to the
original model \cite{hil98d} Because we intend a quantitative
comparison with the microstructure of $\SSS_\EX$ it is necessary to
retain the same level of resolution.  Hence we use throughout this
article the original correlation function $G_\EX(\XX)$ without
subsampling.  Because $G_\EX(r)$ is nearly $0$ for $r>30a$ we have
truncated $G_\EX(r)$ at $r=30a$ to save computer time.  The final
configuration $\SSS_\GF$ with $M=256$ generated by Gaussian filtering
reconstruction that is used in the comparison to experiment is
displayed in Figure \ref{smplC}.

\subsection{Simulated Annealing Reconstruction Model}
The simulated annealing ({\sf SA}) reconstruction model is a second
method to generate a threedimensional random microstructure with
prescribed porosity and correlation function.  A simplified
implementation was recently discussed in Ref. \cite{YT98a} and we
follow their algorithm here.  The method generates a configuration
$\SSS_\SA$ by minimizing the deviations between $G_\SA(\XX)$ and a
predefined reference function $G_0(\XX)$.  Of course in our case we
have again the Fontainebleau sandstone as reference,
i.e. $G_0(\XX)=G_\EX(\XX)$.

The reconstruction is performed on a cubic lattice with side length
$M=M_1=M_2=M_3$ and lattice spacing $a$. The lattice is initialized
randomly with $0$'s and $1$'s such that the volume fraction of $0$'s
equals $\phi(\SSS_\EX)$.  This porosity is preserved throughout the
simulation.  For the sake of numerical efficiency the autocorrelation
function is evaluated in a simplified form using \cite{YT98a}
\begin{eqnarray}
  \
  & & \GT_\SA(r)\left(\GT_\SA(0)- \GT_\SA(0)^2\right)+\GT_\SA(0)^2 = \nonumber \\
  \label{eq:g_r}
  & &=\frac{1}{3M^3} \sum_{\mathbf r} \ch{\MMM}({\mathbf r})
  \left(
    \ch{\MMM}({\mathbf r}+r{\mathbf e}_1) +
    \ch{\MMM}({\mathbf r}+r{\mathbf e}_2) +
    \ch{\MMM}({\mathbf r}+r{\mathbf e}_3) 
  \right) 
\end{eqnarray}
where ${\mathbf e}_i$ are the unit vectors in direction of the
coordinate axes, $r=0,\dots,\frac{M}{2}-1$, and where a tilde
$\widetilde{~}$ is used to indicate the directional restriction.  The
sum $\sum_{\mathbf r}$ runs over all $M^3$ lattice sites $\XX$ with
periodic boundary conditions, i.e. $r_i+r$ is evaluated modulo $M$.

We now perform a simulated annealing algorithm to minimize the
"energy" function
\begin{equation}
  E = \sum_r\left(\GT_\SA(r)-G_\EX(r)\right)^2 ,
\end{equation}
defined as the sum of the squared deviations of $\GT_\SA$ from the
experimental correlation function $G_\EX$.  Each update starts with
the exchange of two pixels, one from pore space, one from matrix
space.  Let $n$ denote the number of the proposed update
step. Introducing an acceptance parameter $T_n$, which may be
interpreted as an $n$-dependent temperature, the proposed
configuration is accepted with probability
\begin{equation}
  p = \min\left(1,\exp\left(-\frac{E_n-E_{n-1}}{T_nE_{n-1}}\right)\right) .
\end{equation}
Here the energy and the correlation function of the configuration is
denoted as $E_n$ and $\GT_{\SA,n}$, respectively.  The evaluation of
$\GT_{\SA,n}$ does not require a complete recalculation. It suffices
to update the correlation function $\GT_{\SA,n-1}$ of the previous
configuration by adding or subtracting those products in
(\ref{eq:g_r}) which changed due to the exchange of pixels.  In case
the proposed move is rejected, the old configuration is restored.

The generation of a configuration with correlation $G_\EX$ is achieved
by lowering $T$. At low $T$ the system approaches a configuration that
minimizes the energy function.  In our simulations we lower $T_n$ with
$n$ as
\begin{equation}
  T_n = \exp\left(-\frac{n}{100000}\right)  .
\end{equation}
We stop the simulation when 20000 consecutive updates are rejected. In
our simulation this happened after $2.5\times 10^8$ updates ($\approx
15$ steps per site).  The resulting configuration $\SSS_\SA$ for the
simulated annealing reconstruction is displayed in Figure \ref{smplD}.

Our definition of the correlation function in (\ref{eq:g_r}) deserves
some comment. A complete evaluation of the correlation function as
defined in (\ref{eq:g_gen}) requires such a great numerical expense
that the algorithm is too slow to allow threedimensional
reconstructions within a reasonable time.  Therefore, to increase the
speed of the algorithm, the correlation function is only evaluated
along the directions of the coordinate axes as indicated in
(\ref{eq:g_r}). As a result of this simplification the reconstructed
sample may cease to be isotropic.  It will in general deviate from the
reference correlation function in all directions other than those of
the axes. In the special case of the correlation function of the
Fontainebleau sandstone, however, this effect seems to be small (see
below).  This may serve as an a posteriori justification for using
(\ref{eq:g_r}).

\section{Results and Discussion}
\label{results}
We begin our presentation of the results with an analysis of
traditional quantities such as porosities and correlation functions of
the four samples.  Then we proceed to a visual characterization of the
threedimensional images.  Next we shall discuss local porosities and
percolation probabilities, and finally we conclude with implications
for transport properties.

\subsection{Conventional Analysis}
Table \ref{ovrvw} gives a synopsis of different properties of the four
samples.  The preparation of the various samples was described in
detail in section \ref{samples}.  The dimensions and porosities also
need no further comment.  Samples \GF~ and \SA~ were constructed to
have the same correlation function as sample \EX.  This is indicated
in the line labeled $G(\XX)$.  In Figure \ref{corrf} we plot the
directionally averaged correlation functions
$G(r)=(G(r,0,0)+G(0,r,0)+G(0,0,r))/3$ of the four samples where
$G(r_1,r_2,r_3)=G(\XX)$.  $G_\DM(r)$ differs clearly from the rest.
Accidentally, however, $G_\DM(0,0,r)\approx G_\EX(0,0,r)$.  $G_\GF(r)$
differs from $G_\EX(r)$ for small $r$ as discussed in section \ref{GF}
above.  Remember also that by construction $G_\GF(r)$ is not expected
to equal $G_\EX(\XX)$ for $r$ larger than 30.  The discrepancy at
small $r$ reflects the quality of the linear filter, and it is also
responsible for the differences of the porosity and specific internal
surface.  Although the reconstruction method of sample $\SSS_\SA$ is
intrinsically anisotropic the correlation function of sample \SA~
agrees also in the diagonal directions with that of sample \EX.
Sample $\SSS_\DM$ on the other hand has an anisotropic correlation
function.

If two samples have the same correlation function they are also
expected to have also the same specific internal surface as calculated
from
\begin{equation}
S = \left.-4\gex{\phi}(1-\gex{\phi})\frac{dG(r)}{dr}\right|_{r=0} .
\end{equation}
The next line in Table \ref{ovrvw} labeled $S$ gives the specific
internal surfaces.

If one defines a decay length by the first zero of the correlation
function then the decay length is roughly $18a$ for samples \EX, \GF~
and \SA.  For sample \DM~ it is somewhat smaller mainly in the $x$-
and $y$-direction.  The correlation length, which will be of the order
of the decay length, is thus relatively large compared to the system
size.  Together with the fact that the percolation threshold for
continuum systems is typically around $0.15$ this might explain why
models
\GF~ and \SA~ are connected in spite of their low value of the porosity.

In summary, the samples $\SSS_\GF$ and $\SSS_\SA$ were constructed to
be indistinguishable with respect to porosity and correlations from
$\SSS_\EX$.  The imperfection of the reconstruction method for sample
\GF~, however, accounts for the deviations of its correlation function
at small $r$ from that of sample \EX.

\subsection{Visual inspection of images}
We now collect results from a visual comparison.  Visual inspection of
Figures \ref{smplA} through \ref{smplD} reveals that none of the
models $\SSS_\DM,\SSS_\GF$ or $\SSS_\SA$ resemble closely the
experimental microstructure $\SSS_\EX$.  This applies in particular to
samples \GF~ and \SA~ which were constructed to match the traditional
geometrical characteristics of sample \EX, such as porosity, specific
surface and correlation function.

Figures \ref{smplA} through \ref{smplD} suggest that samples
$\SSS_\GF$ and $\SSS_\SA$ have isolated islands of matrix space
although this cannot be seen directly because the pore space is
rendered opaque.  Isolated islands of matrix space cannot exist in a
real porous medium such as sample \EX.  They are also absent in the
compaction and diagenesis model \DM.  The comparison is indicated in
the line labeled ``isolated $\MMM$'' in Table \ref{ovrvw}.  The pore
surfaces in samples \GF~ and \SA~ are much rougher than in samples
\EX~ and \DM.  Sample \DM~ appears visually more homogeneous than the
other samples.  Although there is no anisotropy visible for sample
\DM~ from Figure \ref{smplB} its connectivity properties will be found
below to be strongly anisotropic.

In summary the traditional characteristics such as porosity, specific
surface and correlation functions are insufficient to distinguish
different microstructures.  Visual inspection of the pore space by the
human eye indicates that samples \GF~ and \SA~ have a similar
structure which, however, differs from the structure of sample
\EX. Although sample \DM~ resembles sample \EX~ more closely with
respect to surface roughness it differs visibly in the shape of the
grains.

\subsection{Local Porosity Analysis}
We turn to an analysis of the fluctuations in local porosities.  The
differences in visual appearance of the microstructures find a
quantitative expression here.

The local porosity distributions $\mu(\phi,20)$ of the four samples at
$L=20a$ are displayed as the solid lines in Figures
\ref{lppA} through \ref{lppD}.
The ordinates for these curves are plotted on the right vertical axis.
The figures show that the original sample exhibits stronger porosity
fluctuations than the three model samples except for sample \SA~ which
comes close.  Sample \DM~ has the narrowest distribution which
indicates that it is most homogeneous.  Figures \ref{lppA}--\ref{lppD}
show also that the component at the origin, $\mu(0,20)$, is largest
for sample \EX, and smallest for sample \GF.  For samples \DM~ and
\SA~ the values of $\mu(0,20)$ are intermediate and comparable.
Plotting $\mu(0,L)$ as a function of $L$ we find that this remains
true for all $L$.  These results indicate that the experimental sample
\EX~ is more strongly heterogeneous than the models, and that large
regions of matrix space occur more frequently in sample \EX.  A
similar conclusion may be drawn from the variance of local porosity
fluctuations which will be studied below.  The conclusion is also
consistent with the results for $L^*$ shown in Table \ref{ovrvw}.
$L^*$ gives the sidelength of the largest cube that can be fit into
matrix space, and thus $L^*$ may be viewed as a measure for the size
of the largest grain.  Table \ref{ovrvw} shows that the experimental
sample has a larger $L^*$ than all the models.  It is interesting to
note that plotting $\mu(1,L)$ versus $L$ also shows that the curve for
the experimental sample lies above those for the other samples for all
$L$.  Thus, also the size of the largest pore and the pore space
heterogeneity are largest for sample \EX.  If $\mu(\phi,L^*)$ is
plotted for all four samples one finds two groups.  The first group is
formed by samples \EX~ and \DM, the second by samples \GF~ and \SA.
Within each group the curves $\mu(\phi,L^*)$ nearly overlap, but they
differ strongly between them.

Figures \ref{alp}, \ref{vlp}, and \ref{slp} exhibit the dependence of
the local porosity fluctuations on $L$.  In Figure \ref{vlp} we plot
the variance of the local porosity fluctuations, defined in
eq.(\ref{variance}) as function of $L$. The variances for all samples
indicate an approach to a $\delta$-distribution according to
eq. (\ref{lpd5}).  Again sample \DM~ is most homogeneous in the sense
that its variance is smallest.  The agreement between samples \EX,
\GF~ and \SA~ reflects the agreement of their correlation functions,
and is expected by virtue of eq. (\ref{variance}).  Figure \ref{slp}
shows the skewness as a function of $L$ calculated from
\begin{equation}
  \kappa_3(L) = \frac{\overline{(\phi(L)-\phb(L))^3}}{\sigma(L)^3}
\end{equation}
where $\sigma(L)$ is the variance defined in eq. (\ref{variance}).
$\kappa_3$ characterizes the asymmetry of the distribution, and the
difference between the most probable local porosity and its average.
Again samples \GF~ and \SA~ behave similarly, but sample \DM~ and
sample \EX~ differ from each other, and from the rest.

At $L=4a$ the local porosity distributions $\mu(\phi,4)$ show small
spikes at equidistantly spaced porosities for samples \EX~ and \DM,
but not for samples \GF~ and \SA.  The spikes indicate that models
\EX~ and \DM~ have a smoother surface than models \GF~ and \SA. For
smooth surfaces and small measurement cell size porosities
corresponding to an interface intersecting the measurement cell occur
with higher frequency, and this gives rise to spikes.  The presence of
isolated islands of pore or matrix space reduces these spikes.  It is
unclear at present whether the spikes persist when the measurement
cells are chosen to be nonoverlapping.

\subsection{Local Percolation Analysis}
Visual inspection of Figures \ref{smplA} through \ref{smplD} did not
allow us to recognize the degree of connectivity of the various
samples.  A quantitative characterization of the connectivity is
provided by the local percolation probabilities
\cite{hil91d,hil98a}, and
it is here that the samples differ most dramatically.

The samples \EX, \DM~, \GF~ and \SA~ are globally connected in all
three directions.  This, however, does not imply that they have
similar connectivity.  The last line in Table \ref{ovrvw} gives the
fraction of blocking cells at the porosity $0.1355$ and for $L^*$.  It
gives a first indication that the connectivity of samples \DM~ and
\GF~ is, in fact, much poorer than that of the experimental sample
\EX.

Figures \ref{lppA} through \ref{lppD} give a more complete account of
the situation by exhibiting $\lambda_\alpha(\phi,20)$ for
$\alpha=3,c,x,y,z$ for all four samples.  First one notes that sample
\DM~ is strongly anisotropic in its connectivity.  It has a higher
connectivity in the $z$-direction than in the $x$- or $y$-direction.
This might be due to the anisotropic compaction process.
$\lambda_z(\phi,20)$ for sample \DM~ differs from that of sample \EX~
although their correlation functions in the $z$-direction are very
similar.  The $\lambda$-functions for samples \EX~ and \DM~ rise much
more rapidly than those for samples \GF~ and \SA.  The inflection
point of the $\lambda$-curves for samples
\EX~ and \DM~ is much closer to the most probable porosity
(peak) than in samples \GF~ and \SA.  All of this indicates that
connectivity in cells with low porosity is higher for samples \EX~ and
\DM~ than for samples \GF~ and \SA.  In samples \GF~ and \SA~ only
cells with high porosity are percolating on average.  In sample \DM~
the curves $\lambda_x,\lambda_y$ and $\lambda_3$ show strong
fluctuations for $\lambda\approx 1$ at values of $\phi$ much larger
than the $\gex{\phi}$ or $\phi(\SSS_\DM)$.  This indicates a large
number of high porosity cells which are nevertheless blocked.  The
reason for this is perhaps that the linear compaction process in the
underlying model blocks horizontal pore throats and decreases
horizontal spatial continuity more effectively than in the vertical
direction, as shown in
\cite{BO97}, Table 1 p. 142.

The absence of spikes in $\mu(\phi,4)$ for samples \GF~ and \SA~
combined with the fact that cells with average porosity ($\approx
0.135$) are rarely percolating suggests that these samples have a
random morphology similar to percolation.

\subsection{Implications for Transport Properties}
The connectivity analysis of local porosity theory allows to make some
predictions for transport properties (such as conductivity or
permeability) without actually calculating them.  A detailed
comparison between the predictions of local porsity theory and exact
calculation of transport properties will appear elsewhere \cite{WBH99}
These predictions are made by calculating the total fraction of
percolating cells eq. (\ref{pL1}).  The insets in Figures \ref{lppA}
through \ref{lppD} show the functions
$p_\alpha(L)=\overline{\lambda_\alpha(\phi,L)}$ for $\alpha=3,x,y,z,c$
for each sample.  The curves for samples \EX~ and \DM~ are similar but
differ from those for samples \GF~ and \SA.  In Figure \ref{total} we
plot the curves $p_3(L)$ of all four samples in a single figure.  The
samples fall into two groups \{\EX,\DM\} and \{\GF,\SA\} that behave
very differently.  Figure \ref{total} shows that reconstruction
methods
\cite{adl92,YT98a} based on correlation functions
do not reproduce the connectivity properties of porous media.  As a
consequence, within the effective medium approximation of local
porosity theory \cite{hil91d} samples \GF~ and \SA~ would both yield
much lower permeabilities or conductivities than those of samples \EX~
and \DM.  Based on these results it appears questionable whether
correlation function reconstruction can produce reliable models for
the prediction of transport.

\vspace*{3cm}

ACKNOWLEDGEMENT:
The authors are grateful to Dr.David Stern,
(Exxon Reserach Production Company) for providing 
the experimental data set, and to the
Deutsche Forschungsgemeinschaft for financial support.

\newpage
\section*{Table Captions}

\newcounter{tab}
\begin{list}{\textbf{Table \arabic{tab}:}}
{\usecounter{tab}
\setlength{\labelwidth}{2.2cm}
\setlength{\labelsep}{0.3cm}
\setlength{\itemindent}{0pt}
\setlength{\leftmargin}{2.5cm}
\setlength{\rightmargin}{0cm}
\setlength{\parsep}{0.5ex plus0.2ex minus0.1ex}
\setlength{\itemsep}{0ex plus0.2ex}}
\item \label{lpp}
Legend for index $\alpha$ of local percolation
probabilities $\lambda_\alpha(\phi,L)$.
\item \label{ovrvw}
Overview of various properties for the four samples
\end{list}

\newpage
\begin{center}
{\small TABLE \ref{lpp}}\\[12pt]
\begin{tabular}{|c|c|} \hline
index $\alpha$ & meaning \\\hline
$x$ & $x$-direction\\
$y$ & $y$-direction\\
$z$ & $z$-direction\\
$3$ & ($x\wedge y\wedge z$)-direction\\
$c$ & ($x\vee y\vee z$)-direction\\\hline
\end{tabular}
\end{center}

\begin{center}
{\small TABLE \ref{ovrvw} }\\[12pt]
\begin{tabular}{|l||c|c|c|c|} \hline
Properties & $\SSS_\EX$ & $\SSS_\DM$ 
& $\SSS_\GF$ & $\SSS_\SA$\\\hline
Origin & Experiment & Diagenesis Model & Gaussian Field & Simulated
Annealing\\
$M_1\times M_2 \times M_3$ & $300\times 300\times 299$ & 
$255\times 255\times 255$ & $256\times 256\times 256$ & 
$256\times 256\times 256$\\
$\phi(\SSS)$ & 0.1355 & 0.1356 & 0.1421 & 0.1354\\
$G(\XX)$ & $G_\EX$ & $G_\DM$ & $G_\GF\approx G_\EX$ & $G_\SA=G_\EX$\\
$S$ from $\left.\frac{dG}{dr}\right|_{r=0}$ & 0.078  & 0.082 & 0.125 & 0.083\\
Isotropy & $xyz$ & $xy$ & $xyz$ & $xyz$\\
isolated $\MMM$ & No & No & Yes & Yes\\
Pore surface & smooth & smooth & rough & rough \\
$L^*$ & 35 & 25 & 23 & 27\\
Connectivity & $xyz$ & $xyz$ & $xyz$ & $z$\\
$1-\lambda_c(0.1355,L^*)$ & 0.0045 & 0.0239 & 0.3368 & 0.3527\\
\hline
\end{tabular}
\end{center}

\bibliographystyle{ieeetr}

\begin{thebibliography}{10}

\bibitem{fat56a}
I.~Fatt, ``The network model of porous media {I}. capillary pressure
  characteristics,'' {\em AIME Petroleum Transactions}, vol.~207, p.~144, 1956.

\bibitem{sch74}
A.~Scheidegger, {\em The Physics of Flow Through Porous Media}.
\newblock Toronto: University of Toronto Press, 1974.

\bibitem{CD77}
I.~Chatzis and F.~Dullien, ``Modelling pore structure by 2-d and 3-d networks
  with applications to sandstones,'' {\em J. of Canadian Petroleum Technology},
  p.~97, Jan-Mar 1977.

\bibitem{zim82}
J.~Ziman, {\em Models of Disorder}.
\newblock Cambridge: Cambridge University Press, 1982.

\bibitem{RS85}
J.~Roberts and L.~Schwartz, ``Grain consolidation and electrical conductivity
  in porous media,'' {\em Phys. Rev. B}, vol.~31, p.~5990, 1985.

\bibitem{JA87}
C.~Jacquin and P.~Adler, ``Fractal porous media {II}: Geometry of porous
  geological structures,'' {\em Transport in Porous Media}, vol.~2, p.~28,
  1987.

\bibitem{SK87}
L.~Schwartz and S.~Kimminau, ``Analysis of electrical conduction in the grain
  consoliation model,'' {\em Geophysics}, vol.~52, p.~1402, 1987.

\bibitem{oxa91}
U.~Oxaal, ``Fractal viscous fingering in inhomogeneous porous models,'' {\em
  Phys. Rev. A}, vol.~44, p.~5038, 1991.

\bibitem{adl92}
P.~Adler, {\em Porous Media}.
\newblock Boston: Butterworth-Heinemann, 1992.

\bibitem{BT93}
R.~Blumenfeld and S.~Torquato, ``Coarse graining procedure to generate and
  analyze heterogeneous materials: Theory,'' {\em Phys. Rev. E}, vol.~48,
  p.~4492, 1993.

\bibitem{sah95}
M.~Sahimi, {\em Flow and Transport in Porous Media and Fractured Rock}.
\newblock Weinheim: VCH Verlagsgesellschaft mbH, 1995.

\bibitem{FJ95}
J.~Feder and T.~J{\o}ssang, ``Fractal patterns in porous media flow,'' in {\em
  Fractals in Petroleum Geology and Earth Processes} (C.~Barton and P.~L.
  Pointe, eds.), (New York), p.~179, Plenum Press, 1995.

\bibitem{JC96}
D.~Jeulin and A.~L. Co\"{e}nt, ``Morphological modeling of random composites,''
  in {\em Continuum Models and Discrete Systems} (K.~Markov, ed.), (Singapore),
  p.~199, World Scientific Publishing Company, 1996.

\bibitem{AAMHSS97}
J.~Andrade, M.~Almeida, J.~M. Filho, S.~Havlin, B.~Suki, and H.~Stanley,
  ``Fluid flow through porous media: The role of stagnant zones,'' {\em
  Phys.Rev.Lett.}, vol.~79, p.~3901, 1997.

\bibitem{OBA98}
P.~{\O}ren, S.~Bakke, and O.~Arntzen, ``Extending predictive capabilities to
  network models,'' {\em SPE Journal}, 1998.

\bibitem{hil95d}
R.~Hilfer, ``Transport and relaxation phenomena in porous media,'' {\em
  Advances in Chemical Physics}, vol.~XCII, p.~299, 1996.

\bibitem{hil96g}
C.~Andraud, A.~Beghdadi, E.~Haslund, R.~Hilfer, J.~Lafait, and B.~Virgin,
  ``Local entropy characterization of correlated random microstructures,'' {\em
  Physica A}, vol.~235, p.~307, 1997.

\bibitem{hil92f}
F.~Boger, J.~Feder, R.~Hilfer, and T.~J{\o}ssang, ``Microstructural sensitivity
  of local porosity distributions,'' {\em Physica A}, vol.~187, p.~55, 1992.

\bibitem{sic97}
C.~van Siclen, ``Information entropy of complex structures,'' {\em Phys. Rev.
  E}, vol.~56, p.~5211, 1997.

\bibitem{qui84}
J.~Quiblier, ``A new three dimensional modeling technique for studying porous
  media,'' {\em J. Colloid Interface Sci.}, vol.~98, p.~84, 1984.

\bibitem{YT98a}
C.~Yeong and S.~Torquato, ``Reconstructing random media,'' {\em Phys.Rev. E},
  vol.~57, p.~495, 1998.

\bibitem{YT98b}
C.~Yeong and S.~Torquato, ``Reconstructing random media {II}. three-dimensional
  media from two-dimensional cuts,'' {\em Phys.Rev. E}, vol.~58, p.~224, 1998.

\bibitem{BZ85}
T.~Bourbie and B.~Zinszner, ``Hydraulic and acoustic properties as a function
  of porosity in {F}ontainebleau snadstone,'' {\em J.Geophys.Res.}, vol.~90,
  p.~11524, 1995.

\bibitem{BCZ87}
T.~Bourbie, O.~Coussy, and B.~Zinszner, {\em Acoustics of Porous Media}.
\newblock Paris: Editions Technip, 1987.

\bibitem{TSA93}
J.~Thovert, J.~Salles, and P.~Adler, ``Computerized chracterization of the
  geometry of real porous media: Their discretization, analysis and
  interpretation,'' {\em J. Microscopy}, vol.~170, p.~65, 1993.

\bibitem{ste85}
G.~Stell, ``Mayer-montroll equations (and some variants) through history for
  fun and profit,'' in {\em The Wonderful World of Stochastics} (M.~Shlesinger
  and G.~Weiss, eds.), (Amsterdam), p.~127, Elsevier, 1985.

\bibitem{dul92}
F.~Dullien, {\em Porous Media - Fluid Transport and Pore Structure}.
\newblock San Diego: Academic Press, 1992.

\bibitem{BO97}
S.~Bakke and P.~{\O}ren, ``3-d pore-scale modeling of sandstones and flow
  simulations in pore networks,'' {\em SPE Journal}, vol.~2, p.~136, 1997.

\bibitem{hil92a}
R.~Hilfer, ``Local porosity theory for flow in porous media,'' {\em Phys. Rev.
  B}, vol.~45, p.~7115, 1992.

\bibitem{hil91d}
R.~Hilfer, ``Geometric and dielectric characterization of porous media,'' {\em
  Phys. Rev. B}, vol.~44, p.~60, 1991.

\bibitem{hil92b}
R.~Hilfer, ``Geometry, dielectric response and scaling in porous media,'' {\em
  Physica Scripta}, vol.~T44, p.~51, 1992.

\bibitem{hil93b}
R.~Hilfer, ``Local porosity theory for electrical and hydrodynamical transport
  through porous media,'' {\em Physica A}, vol.~194, p.~406, 1993.

\bibitem{hil93c}
B.~Hansen, E.~Haslund, R.~Hilfer, and B.~N{\o}st, ``Dielectric dispersion
  measurements of salt water saturated porous glass compared with local
  porosity theory,'' {\em Mater.Res.Soc.Proc.}, vol.~290, p.~185, 1993.

\bibitem{hil94b}
R.~Hilfer, B.N{\o}st, E.Haslund, Th.Kautzsch, B.Virgin, and B.D.Hansen, ``Local
  porosity theory for the frequency dependent dielectric function of porous
  rocks and polymer blends,'' {\em Physica A}, vol.~207, p.~19, 1994.

\bibitem{hil98a}
B.~Biswal, C.~Manwart, and R.~Hilfer, ``Threedimensional local porosity
  analysis of porous media,'' {\em Physica A}, vol.~255, p.~221, 1998.

\bibitem{AJQ90}
P.~Adler, C.~Jacquin, and J.~Quiblier, ``Flow in simulated porous media,'' {\em
  Int.J.Multiphase Flow}, vol.~16, p.~691, 1990.

\bibitem{YFKTA93}
J.~Yao, P.~Frykman, F.~Kalaydjian, P.~Thovert, and P.~Adler, ``High-order
  moments of the phase function for real and reconstructed model porous media:
  A comparison,'' {\em J. of Colloid and Interface Science}, vol.~156, p.~478,
  1993.

\bibitem{hil98d}
B.~Biswal and R.~Hilfer, ``Microstructure analysis of reconstructed porous
  media,'' {\em Physica A}, vol.~266, p.~307, 1999.

\bibitem{WBH99}
J.~Widjajakusuma, B.~Biswal, and R.~Hilfer, ``Quantitative prediction of
  effective material properties of heterogeneous media,'' {\em Comp.Mat.Sci.},
  p.~to appear, 1999.

\end{thebibliography}
\textheight24.5cm

\newpage
\section*{Figure Captions}

\newcounter{fig}
\begin{list}{\textbf{Figure \arabic{fig}:}}
{\usecounter{fig}
\setlength{\labelwidth}{2.2cm}
\setlength{\labelsep}{0.3cm}
\setlength{\itemindent}{0pt}
\setlength{\leftmargin}{2.5cm}
\setlength{\rightmargin}{0cm}
\setlength{\parsep}{0.5ex plus0.2ex minus0.1ex}
\setlength{\itemsep}{0ex plus0.2ex}}
\item \label{smplA}
Threedimensional pore space of
Fontainebleau  sandstone (sample \EX).
The resolution of the image is $a=7.5\mu$m,
the sample dimensions are
$M_1 = 300$,
$M_2 = 300$,
$M_3 = 299$.
The porosity is $\phi(\SSS_\EX)=0.1355$.
The pore space is indicated opaque, the matrix space is
transparent.
The lower image shows the front plane of the sample
as a twodimensional thin section (pore space black,
matrix grey).
\item\label{smplB}
Threedimensional pore space of the sedimentation and
diagenesis model (sample \DM).
The resolution is $a=7.5\mu$m, the sample dimensions are
$M_1 = 255$,
$M_2 = 255$,
$M_3 = 255$.
The bulk porosity is $\phi(\SSS_\DM)=0.1356$.
The pore space is indicated opaque, the matrix space is
transparent.
The lower image shows the front plane of the sample
as a twodimensional thin section (pore space black,
matrix grey).
\item\label{smplC}
Threedimensional pore space having the same correlation
function as the experimental sample of Fontainebleau
sandstone (sample \GF).
The pore space was constructed using Gaussian random
fields which are subsequently filtered.
The resolution is $a=7.5\mu$m, the sample dimensions are
$M_1 = 256$,
$M_2 = 256$,
$M_3 = 256$.
The bulk porosity is $\phi(\SSS_\GF)=0.1421$.
The pore space is indicated opaque, the matrix space is
transparent.
The lower image shows the front plane of the sample
as a twodimensional thin section (pore space black,
matrix grey).
\item\label{smplD}
Threedimensional pore space having the same correlation
function as the experimental sample of Fontainebleau
sandstone (sample \SA).
The pore space was constructed using a simulated annealing
algorithm.
The resolution is $a=7.5\mu$m, the sample dimensions are
$M_1 = 256$,
$M_2 = 256$,
$M_3 = 256$.
The bulk porosity is $\phi(\SSS_\SA)=0.1354$.
The pore space is indicated opaque, the matrix space is
transparent.
The lower image shows the front plane of the sample
as a twodimensional thin section (pore space black,
matrix grey).
\item\label{corrf}
Averaged directional correlation functions 
$G(r)=(G(r,0,0)+G(0,r,0)+G(0,0,r))/3$
of all four samples.
\item\label{lppA}
Local percolation probabilities $\lambda_\alpha(\phi,20)$
(broken curves, values on left axis) 
and local porosity distribution $\mu(\phi,20)$ 
(solid curve, values on right axis) at $L=20$
for sample \EX.
The inset shows the function $p_\alpha(L)$.
The line styles correponding to $\alpha=c,x,y,z,3$
are indicated in the legend.
\item\label{lppB}
Local percolation probabilities $\lambda_\alpha(\phi,20)$
(broken curves, values on left axis) 
and local porosity distribution $\mu(\phi,20)$ 
(solid curve, values on right axis) at $L=20$
for sample \DM.
The inset shows the function $p_\alpha(L)$.
The line styles correponding to $\alpha=c,x,y,z,3$
are indicated in the legend.
\item\label{lppC}
Local percolation probabilities $\lambda_\alpha(\phi,20)$
(broken curves, values on left axis) 
and local porosity distribution $\mu(\phi,20)$ 
(solid curve, values on right axis) at $L=20$
for sample \GF.
The inset shows the function $p_\alpha(L)$.
The line styles correponding to $\alpha=c,x,y,z,3$
are indicated in the legend.
\item\label{lppD}
Local percolation probabilities $\lambda_\alpha(\phi,20)$
(broken curves, values on left axis) 
and local porosity distribution $\mu(\phi,20)$ 
(solid curve, values on right axis) at $L=20$
for sample \SA.
The inset shows the function $p_\alpha(L)$.
The line styles correponding to $\alpha=c,x,y,z,3$
are indicated in the legend.
\item\label{alp}
Average local porosities for sample 
\EX (solid line with tick)
\DM (dashed line with cross)
\GF (dotted line with square), and 
\SA (dash-dotted line with circle).
\item\label{vlp}
Variance of local porosities for sample
\EX (solid line with tick)
\DM (dashed line with cross)
\GF (dotted line with square), and 
\SA (dash-dotted line with circle).
\item\label{slp}
Skewness of local porosities for sample
\EX (solid line with tick)
\DM (dashed line with cross)
\GF (dotted line with square), and 
\SA (dash-dotted line with circle).
\item\label{total}
$p_3(L)$ for sample
\EX (solid line with tick)
\DM (dashed line with cross)
\GF (dotted line with square), and 
\SA (dash-dotted line with circle).
\end{list}





\newpage
{\small\sffamily B.Biswal et al. \hfill Figure \ref{corrf}}\\[3cm]
\psfrag{r}{\Large $r$}
\psfrag{gr}{\Large $G(r)$}
\epsfig{figure=\DIRECTORY/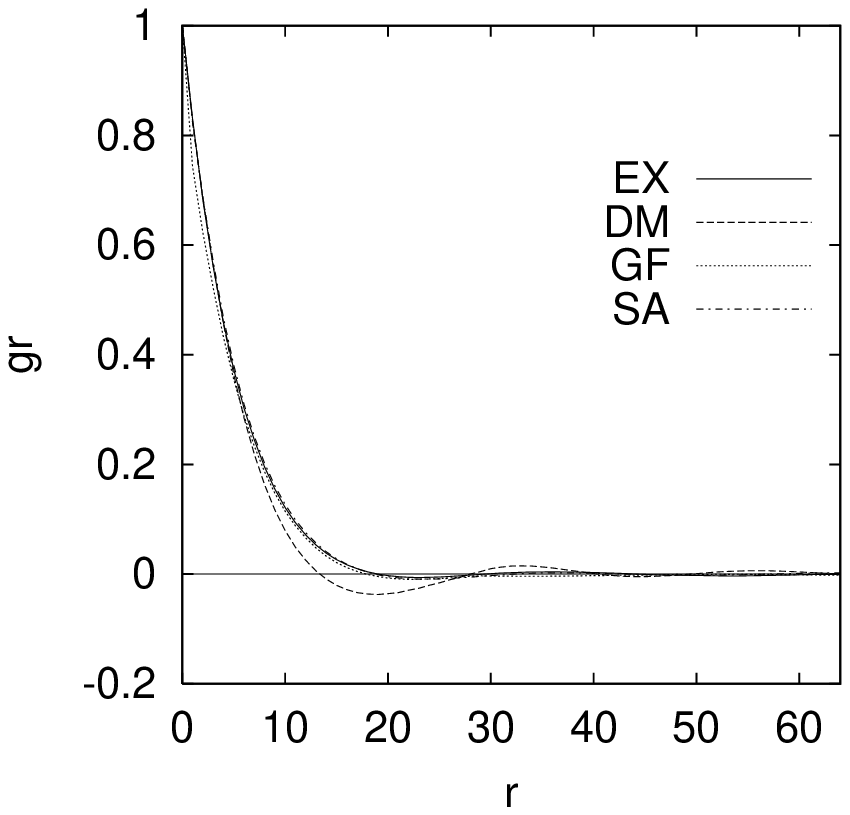,angle=0,width=16cm}

\newpage
{\small\sffamily B.Biswal et al. \hfill Figure \ref{lppA}}\\[3cm]
\psfrag{xlabel}{\Large $\phi$}
\psfrag{ylabel}{\Large $\lambda_\alpha(\phi;20)$}
\psfrag{y2label}{\Large $\mu(\phi;20)$}
\psfrag{p(L)}{ $p_\alpha(L)$}
\psfrag{L}{ $L$}
\psfrag{x1}{0}
\psfrag{x2}{20}
\psfrag{x3}{40}
\psfrag{x4}{60}
\psfrag{y1}{0.0}
\psfrag{y2}{.25}
\psfrag{y3}{.50}
\psfrag{y4}{.75}
\psfrag{y5}{1.0}
\epsfig{figure=\DIRECTORY/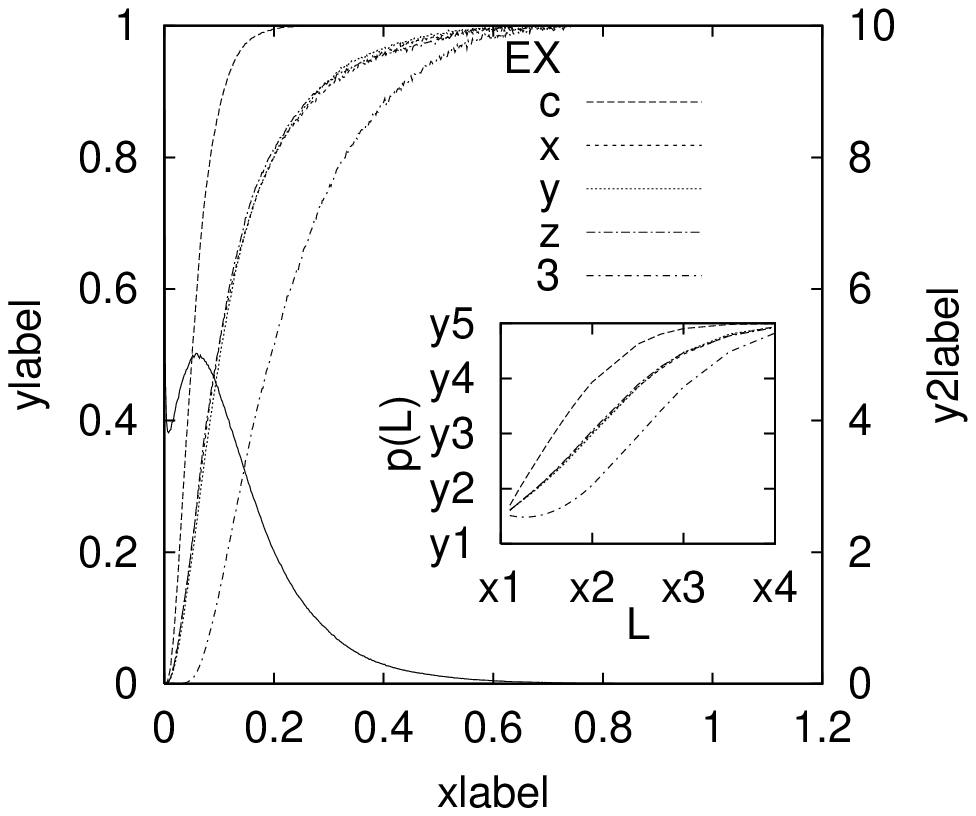,
angle=0,width=16cm}

\newpage
{\small\sffamily B.Biswal et al. \hfill Figure \ref{lppB}}\\[3cm]
\epsfig{figure=\DIRECTORY/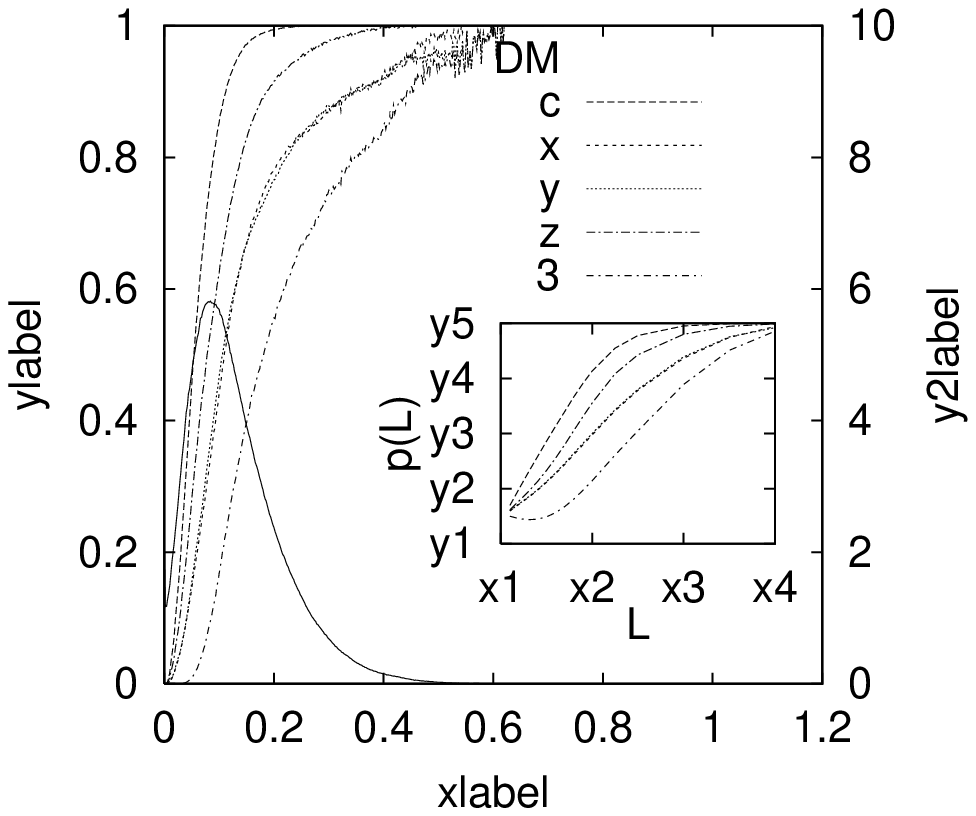,
angle=0,width=16cm}

\newpage
{\small\sffamily B.Biswal et al. \hfill Figure \ref{lppC}}\\[3cm]
\epsfig{figure=\DIRECTORY/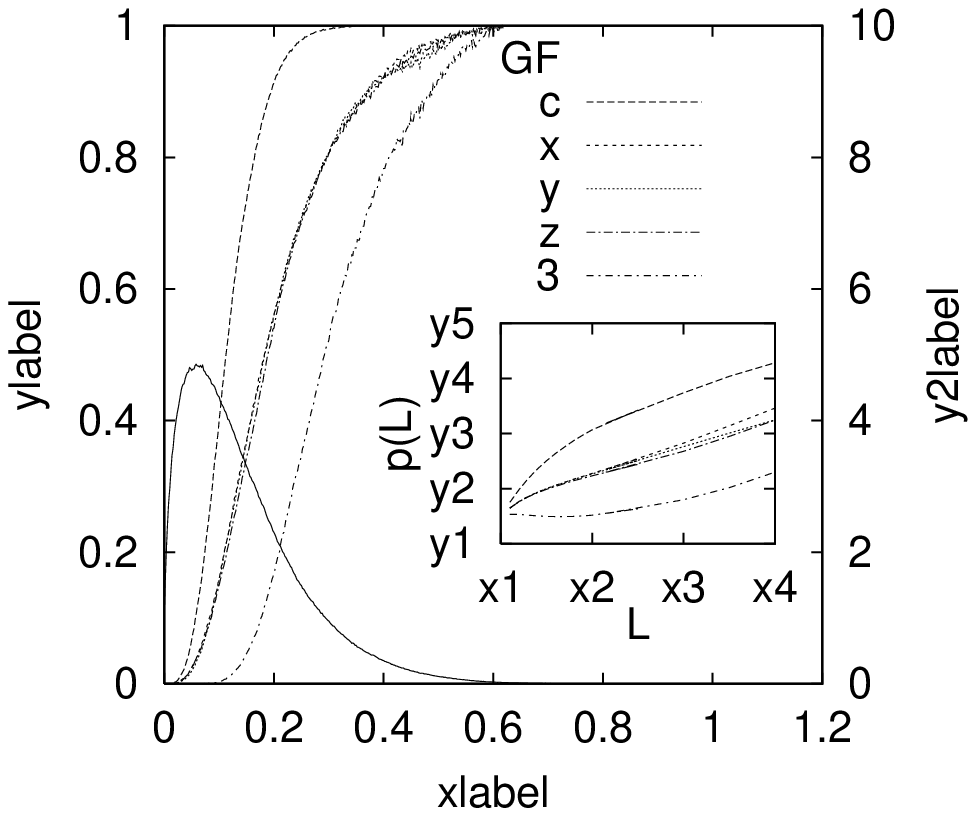,
angle=0,width=16cm}

\newpage
{\small\sffamily B.Biswal et al. \hfill Figure \ref{lppD}}\\[3cm]
\epsfig{figure=\DIRECTORY/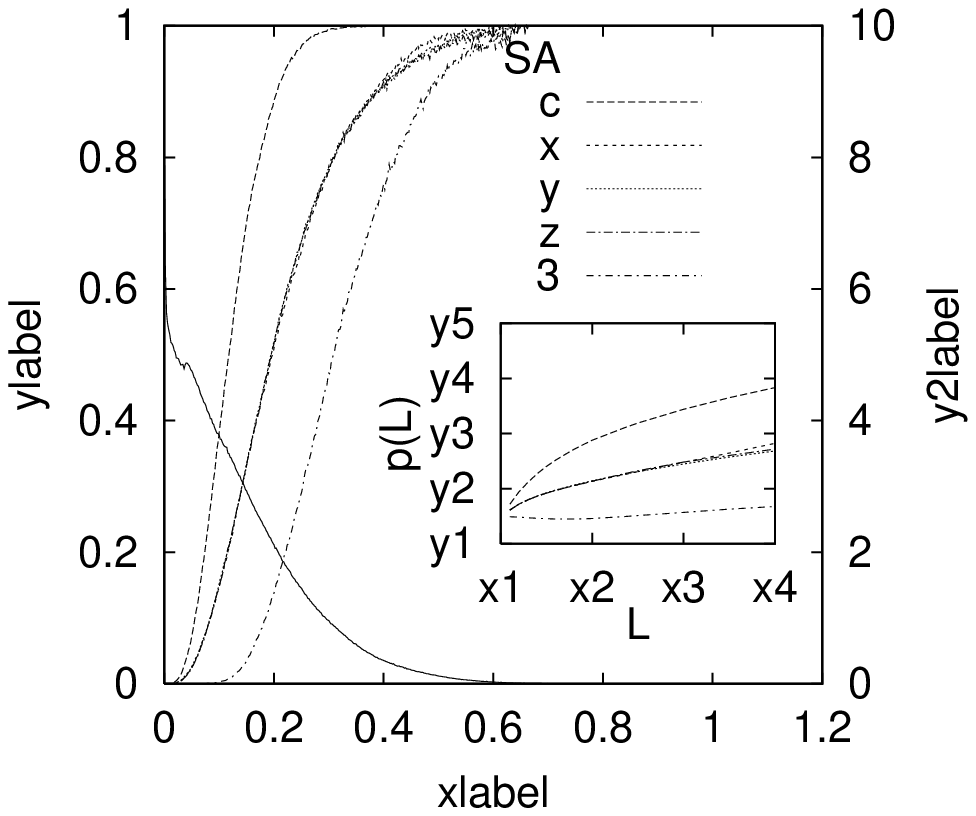,
angle=0,width=16cm}

\newpage
{\small\sffamily B.Biswal et al. \hfill Figure \ref{alp}}\\[3cm]
\psfrag{barphi}{\Large $\bar\phi(L)$}
\epsfig{figure=\DIRECTORY/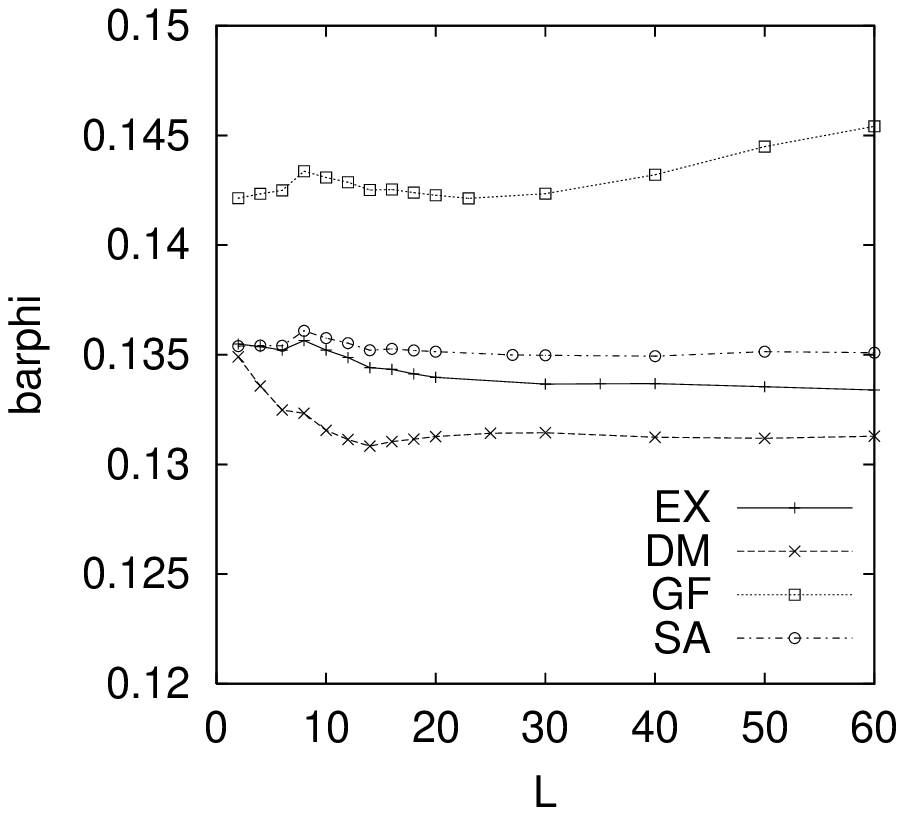,
angle=0,width=16cm}

\newpage
{\small\sffamily B.Biswal et al. \hfill Figure \ref{vlp}}\\[3cm]
\psfrag{sigma}{\Large $\sigma$}
\epsfig{figure=\DIRECTORY/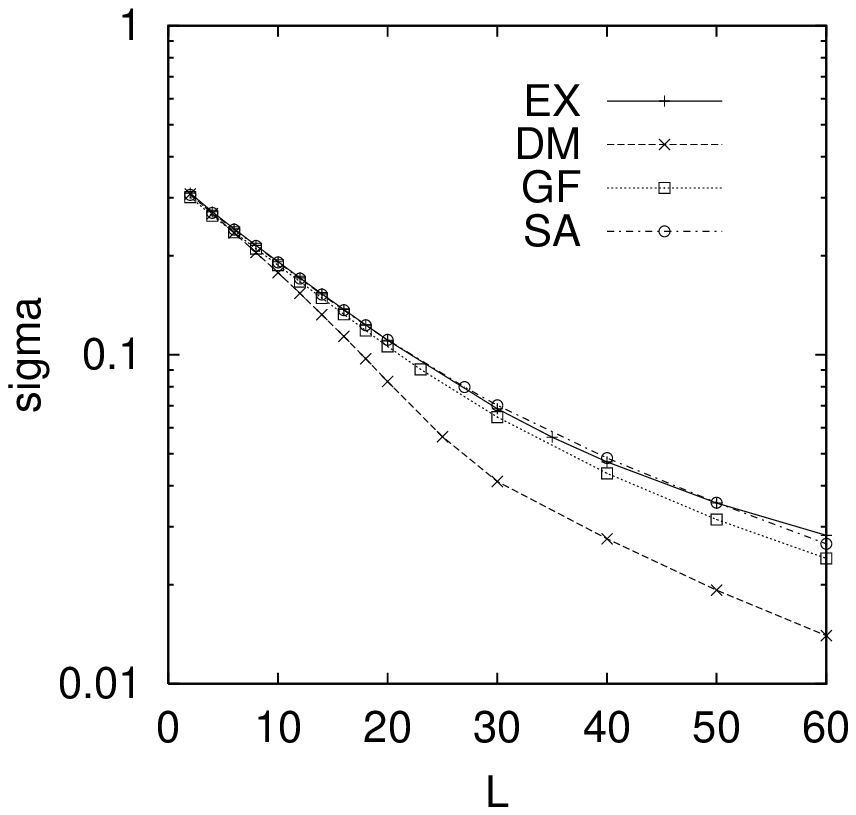,
angle=0,width=16cm}

\newpage
{\small\sffamily B.Biswal et al. \hfill Figure \ref{slp}}\\[3cm]
\psfrag{kappa}{\Large $\kappa_3$}
\epsfig{figure=\DIRECTORY/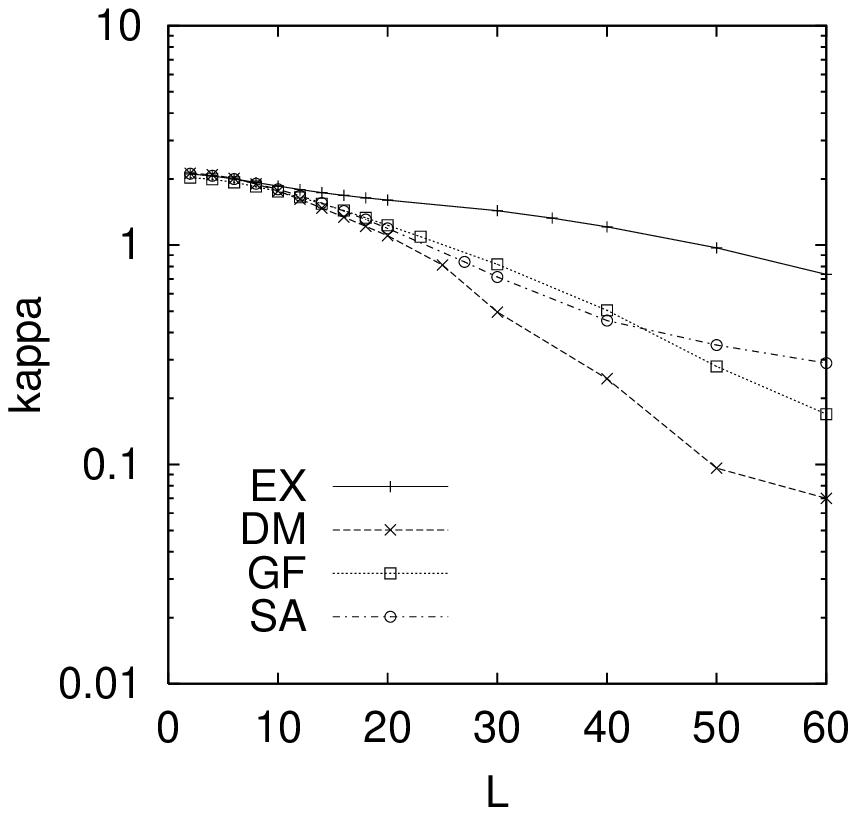,
angle=0,width=16cm}

\newpage
{\small\sffamily B.Biswal et al. \hfill Figure \ref{total}}\\[3cm]
\psfrag{pl3}{\Large $p_3(L)$}
\epsfig{figure=\DIRECTORY/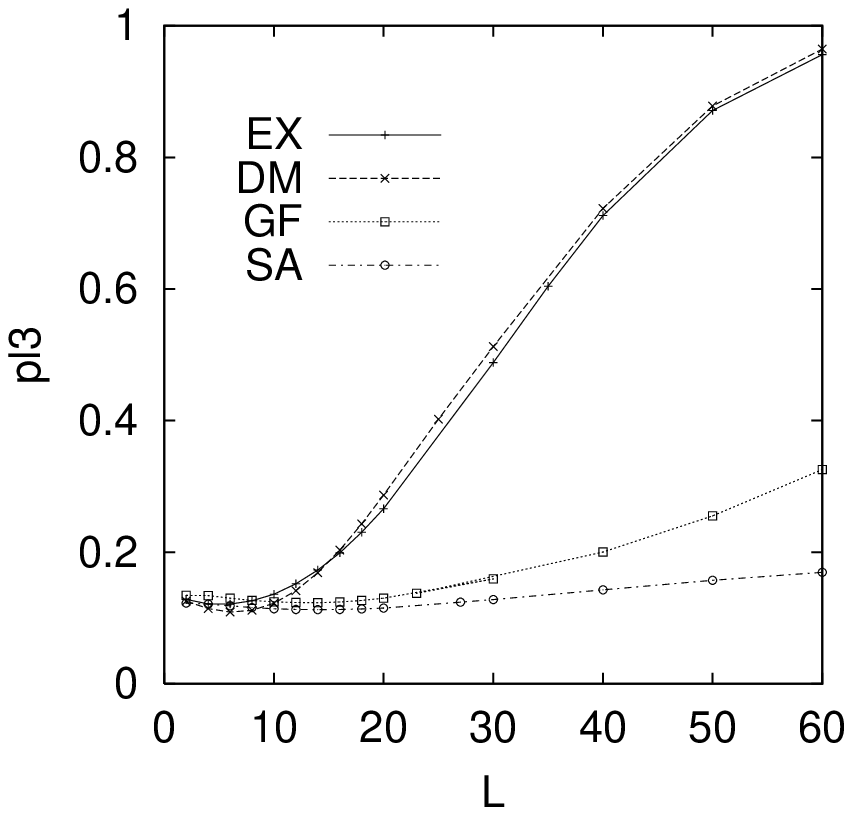,
angle=0,width=16cm}

\voffset0cm

\end{document}